\documentclass{PoS}

\graphicspath{{/home/Schafer/TeX/TALKS/HSQCD2014/new_figures/},
{/home/Schafer/TeX/TALKS/MESON2012/},
{/home/Schafer/TeX/TALKS/UJ_Seminarium/}
{/home/Schafer/TeX/TALKS/Diffraction2014/DIS_2014/}
{/home/Schafer/TeX/TALKS/Gribov_85/}
{/home/Schafer/TeX/TALKS/Gribov_85/figures/}
}

\newcommand{\bp}{{\bf{p}}}
\newcommand{\bq}{{\bf{q}}}

\title{Photoproduction of $J/\psi$ and $\Upsilon$ states in exclusive and proton-dissociative diffractive events}

\ShortTitle{Photoproduction of $J/\psi$ and $\Upsilon$ states in exclusive and proton-dissociative diffractive events}

\author{\speaker{Wolfgang Sch\"afer}%
        \thanks{
       Work reported here partially supported by the Polish National Science 
       Centre grant DEC-2014/15/B/ST2/02528.
        }\\
       Institute of Nuclear Physics, Krakow\\
       E-mail: \email{wolfgang.schafer@ifj.edu.pl}}

\author{Antoni Szczurek\\%
       Institute of Nuclear Physics, Krakow\\
       E-mail: \email{Antoni.Szczurek@ifj.edu.pl}}
       
\author{Anna Cisek\\%
       University of Rzesz\'ow, Rzesz\'ow\\
       E-mail: \email{acisek@ur.edu.pl}}

\abstract{
The amplitude for $\gamma p \to V p$, where $V$ is a $J/\psi$ or $\Upsilon$ ground state or excited 
vector meson,  is calculated in a pQCD $k_{T}$-factorization approach. 
We use this amplitude to predict 
the cross section for exclusive photoproduction of $J/\psi, \psi', \Upsilon$ mesons in proton-proton collisions. 
Calculations are performed for a variety of unintegrated gluon distributions, and we compare our results 
to LHCb data.
Here we especially focus on the possibilty of constraining saturation effects.
Compared to earlier calculations we include both Dirac and Pauli electromagnetic form factors.
We discuss the role of the $Q\bar Q$ light-cone wave
functions for differential distributions for ratios
such as $\sigma(\psi')/\sigma(J/\psi)$.
Absorption effects are always taken into account.

We also discuss the related diffractive production in proton dissociative events.
Here we concentrate on electromagnetic dissociation , which is calculable without additional free parameters. 
Besides being of interest in their own right, dissociative events constitute an important experimental background to exclusive production.
}

\FullConference{XXIV International Workshop on Deep-Inelastic Scattering and Related Subjects\\
		11-15 April, 2016\\
		DESY Hamburg, Germany}

\begin{document}

\section{Introduction}

The flux of Weizs\"acker-Williams photons, which accompany high energy charged particles 
opens a possibility to study high-energy photon-hadron interactions at previously unaccessible energies
in collisions at the LHC \cite{Klein:2003vd}.
Of special interest is the exclusive photoproduction of vector mesons, $pp \to ppV$, and here 
in turn the heavy quarkonia 
play a special role. A rich body of experimental data from $ep$ collisions taken at HERA, 
and a large amount of theoretical works, have convincingly established the 
diffractive vector meson production as a testing ground of the transition
between soft and hard QCD Pomeron physics (for a review, see \cite{Ivanov:2004ax}).

There are a number of crucial differences between the $ep$ and $pp$ collisions. 
Firstly, due to the electromagnetic form factors (finite size) of the proton, virtualities
of photons are strongly bounded (photons are quasireal) in the exclusive $pp \to Vpp$ processes. 
This means that photon virtuality $Q^2$ cannot serve as a hard scale. In order to study the
perturbative QCD (pQCD) driven hard Pomeron, we have to restrict ourselves to heavy quarkonia.
Secondly, both of the incoming protons can be a source of photons, so that two amplitudes
have to be added coherently.
Thirdly, protons are strongly interacting, and the many open inelastic channels require
the evaluation of absorptive corrections to the naive $\gamma-$Pomeron fusion amplitude.
For a discussion of these issues and explicit formulas, see e.g. \cite{Schafer:2007mm,Cisek:2014ala}. 

\section{Exclusive photoproduction of $J/\psi, \psi(2S), \Upsilon$}

The crucial input for the evaluation of the $pp \to ppV$ cross section is the 
photoproduction amplitude for the $\gamma p \to Vp$ subprocess. 
We concentrate on heavy quarkonia, so that the Pomeron exchange can be modelled by a
pQCD gluon ladder.
We write the full amplitude at finite transverse momentum transfer $\Delta_\perp$ as: 
\begin{eqnarray}
{\cal M}(W,\Delta_\perp^2) = (i + \rho) \, \Im m {\cal M}(W,\Delta_\perp^2=0) \cdot f(-\Delta_\perp^2,W) \, . 
\label{full_amp}
\end{eqnarray}
Here, the imaginary amplitude of the forward amplitude can be calculated in terms of the unintegrated 
gluon distribution in the target proton and the light-cone wave function of the vector meson.
For the formalism used, see \cite{Cisek:2014ala}.

The real part of the amplitude is restored from analyticity,
\begin{eqnarray}
\rho = {\Re e {\cal M} \over \Im m {\cal M}} =  
\tan \Big ( {\pi \over 2} \, { \partial \log \Big( \Im m {\cal M}/W^2 \Big) \over \partial \log W^2 } \Big) \, .
\nonumber
\end{eqnarray}
The dependence on momentum transfer $t = -\Delta_\perp^2$ is parametrized by the function
$f(t,W)$, which dependence on energy derives from the Regge slope
\begin{eqnarray}
B(W) = b_0 + 2 \alpha'_{eff} \log \Big( {W^2 \over W^2_0} \Big) \, , 
\end{eqnarray}
where we use: $b_0 = 4.88$, $\alpha'_{eff} = 0.164$ GeV$^{-2}$ and $W_0 = 90$ GeV.

\begin{figure}
\includegraphics[width=.5\textwidth]{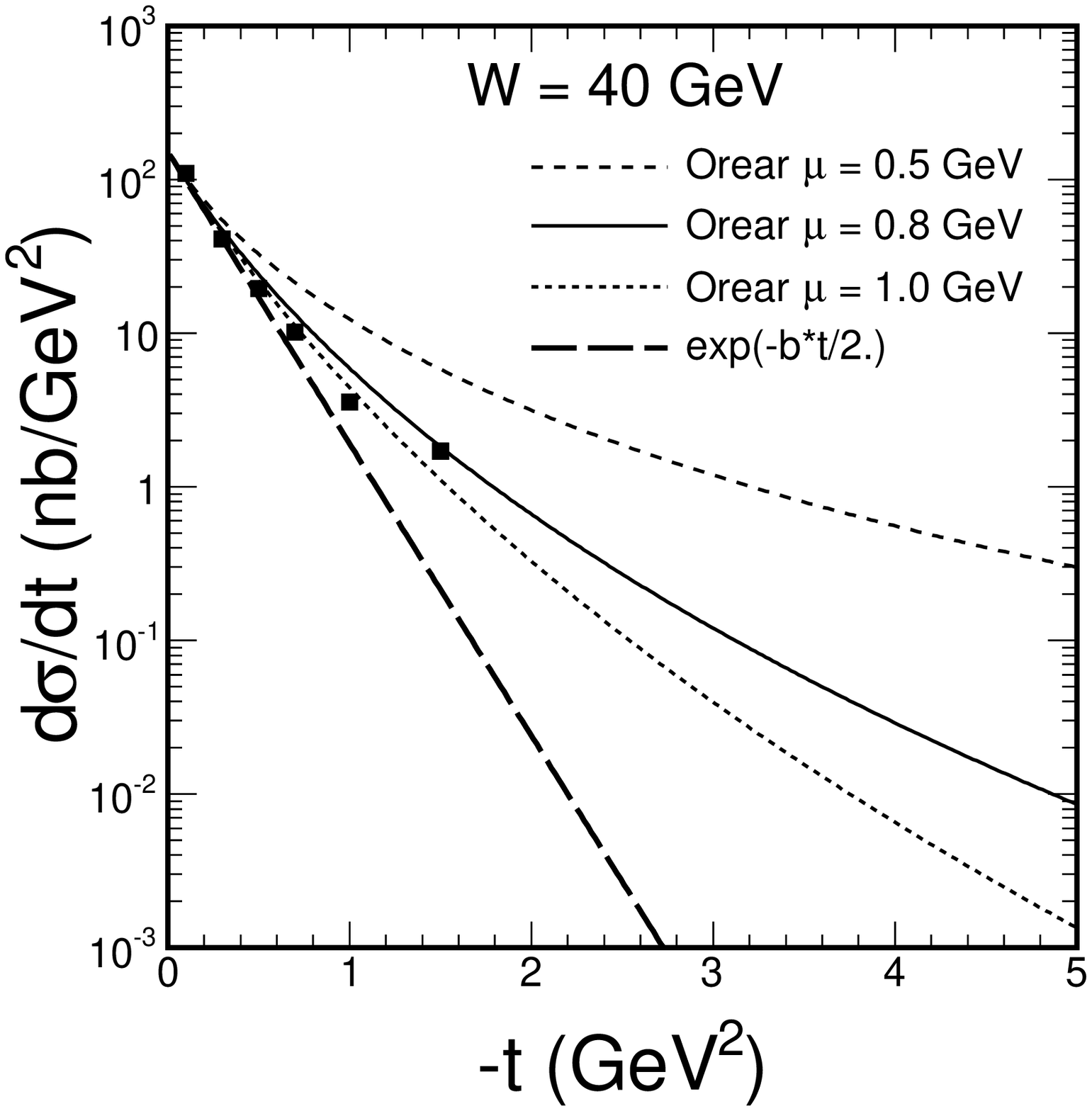}
\includegraphics[width=.5\textwidth]{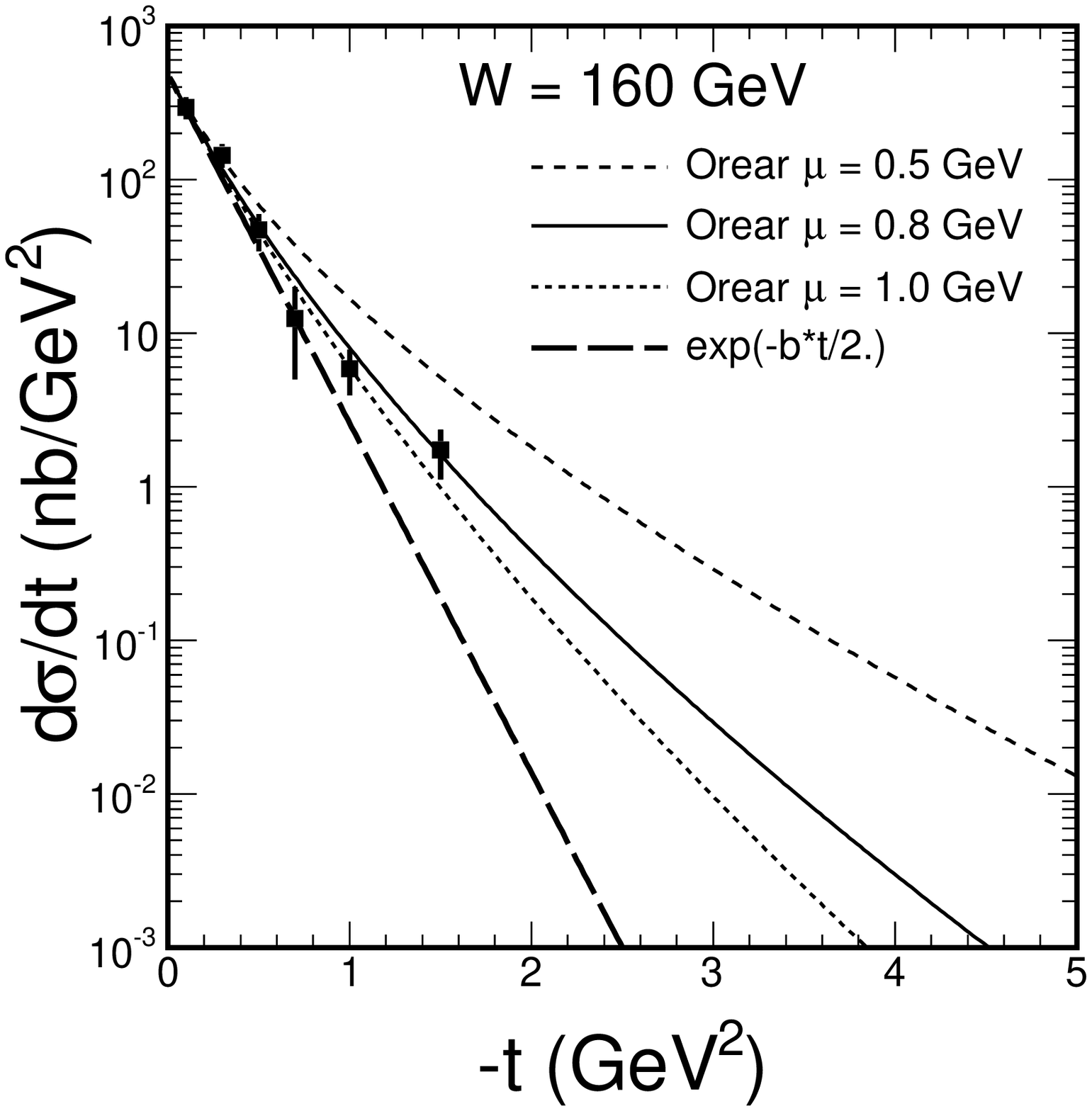}
\caption{
$d\sigma/dt$ for $\gamma p \to J/\psi p$  
for different energies indicated in the figure caption. Data from \cite{Chekanov:2002xi}.
``Orear'' denotes the stretched-exponential parametrization.
}
\label{fig:1}
\end{figure}

Within the diffraction cone it is customary to use the exponential parametrization:
\begin{eqnarray}
 f(t,W) = \exp\Big({1\over 2}B(W) t\Big )  \, .
\label{exponential} 
\end{eqnarray}
For our purposes however an extension to larger $|t| \sim 1 \div 2 \, \rm{GeV}^2$
is desirable. A comparison with ZEUS data \cite{Chekanov:2002xi} (see Fig. \ref{fig:1}) shows that a
``stretched exponential'' parametrization
\begin{eqnarray}
f(t,W) =  \exp ( \mu^2 B(W) ) \exp \Big( - \mu^2 B(W) \sqrt{1 - t/\mu^2} \Big) \; ,
\label{Orear} 
\end{eqnarray}
does a good job with only one new parameter.
What is the physics of the ``hard tail'' encoded in the stretched exponetial?
At this moment we cannot tell, it might as well be a contamination of data by dissociative
events, an effect of multiple scatterings, or -- somewhat unlikely -- ``hard'' scattering.

\begin{figure}
\includegraphics[width=4.8cm]{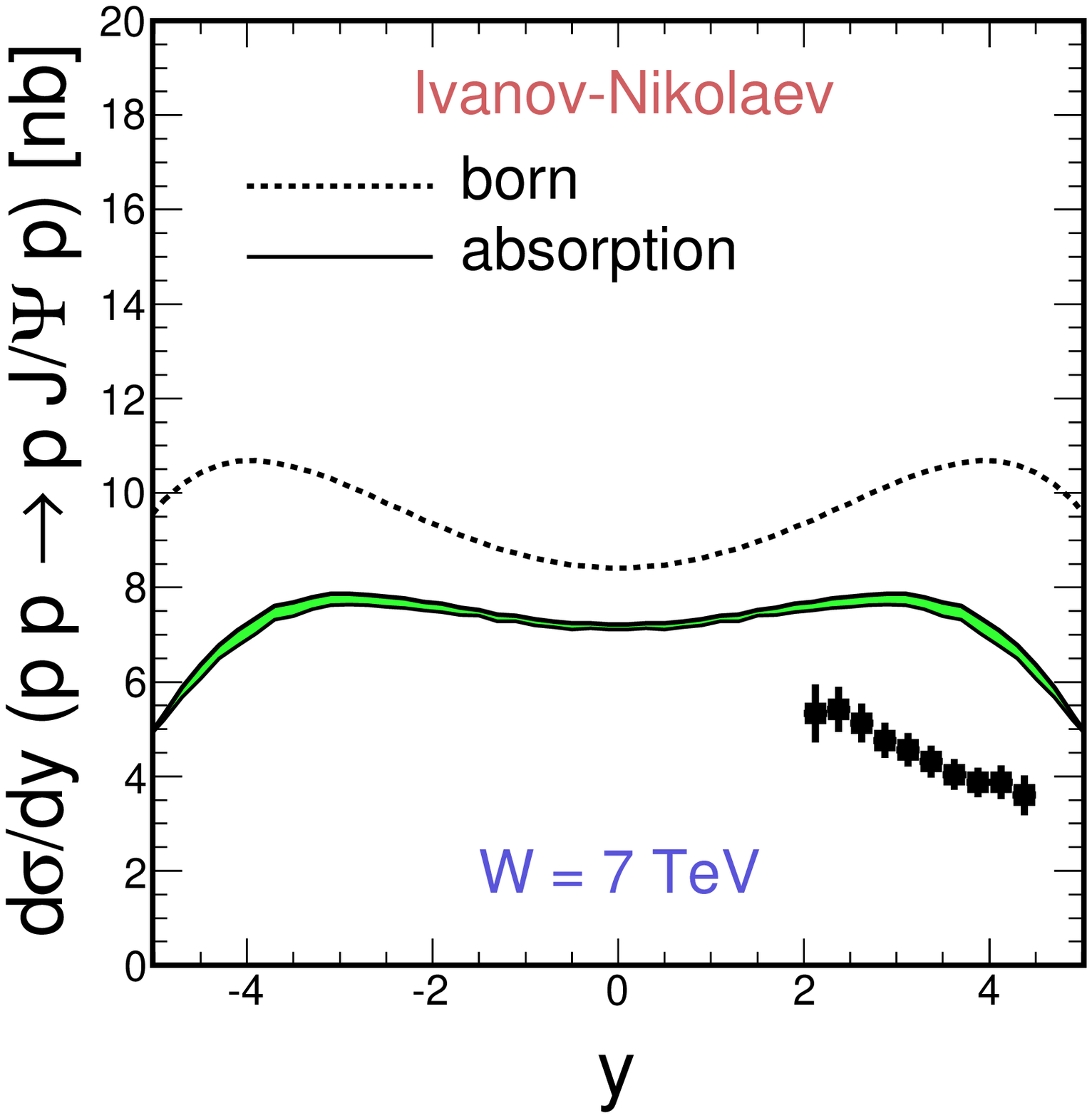}
\includegraphics[width=4.8cm]{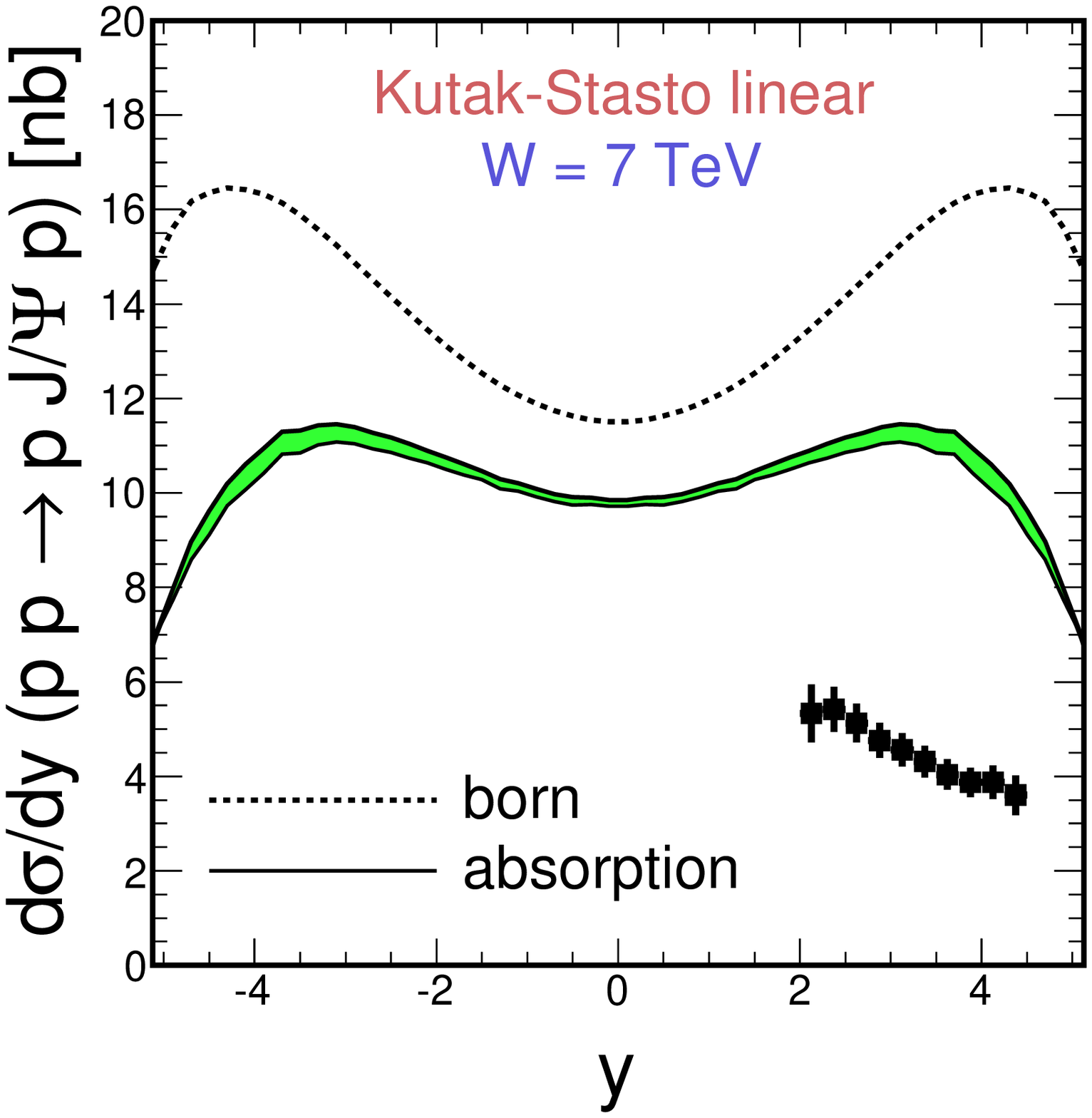}
\includegraphics[width=4.8cm]{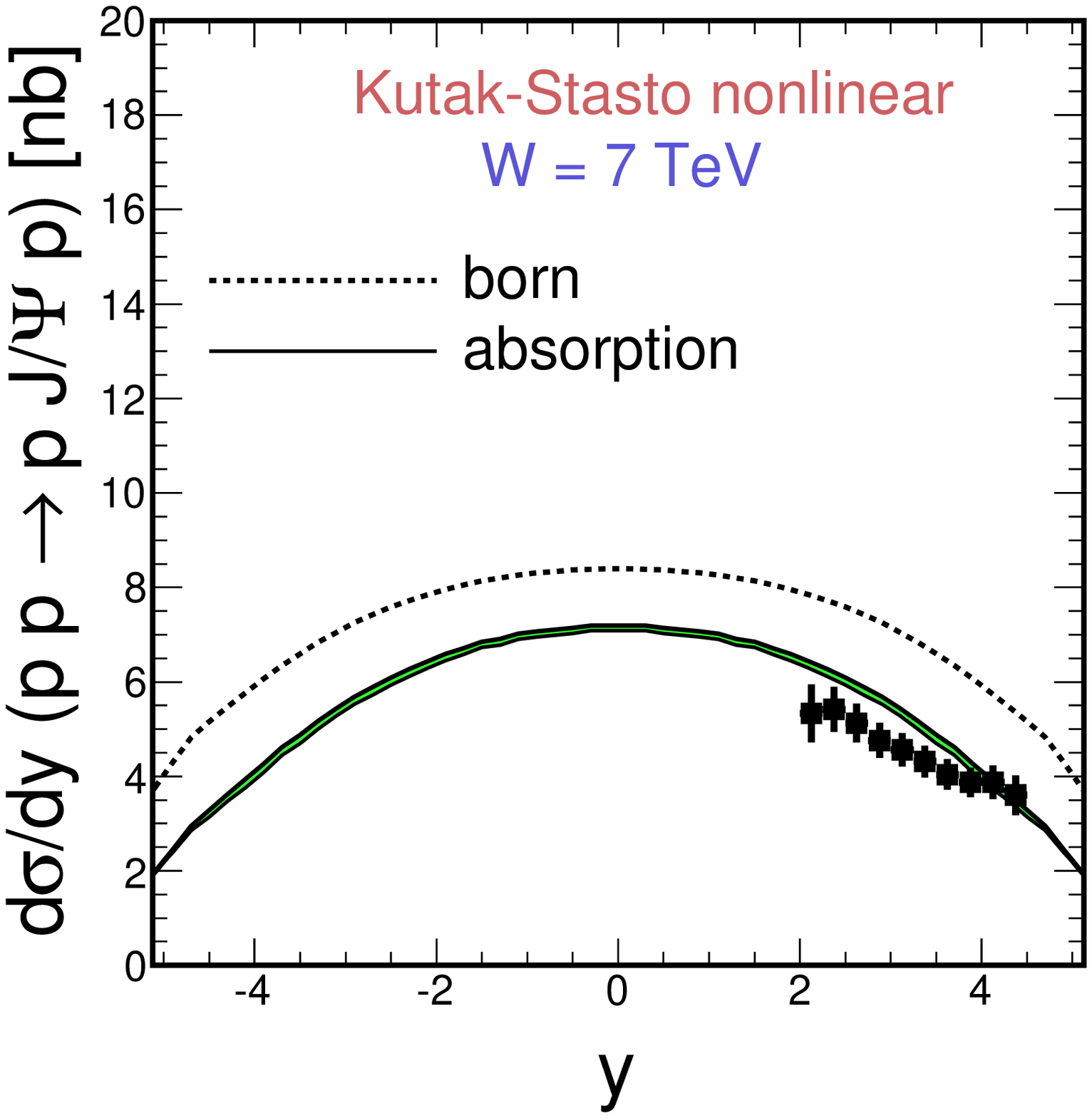}
\caption{ 
$J/\psi$ rapidity distribution calculated with inclusion of absorption effects (solid lines), compared with the Born result (dashed line) for
$\sqrt{s}$ = 7 TeV. 
The LHCb data points \cite{Aaij:2014iea} are shown for comparison.
}
\label{fig:2}
\end{figure}

\begin{figure}
\includegraphics[width=4.8cm]{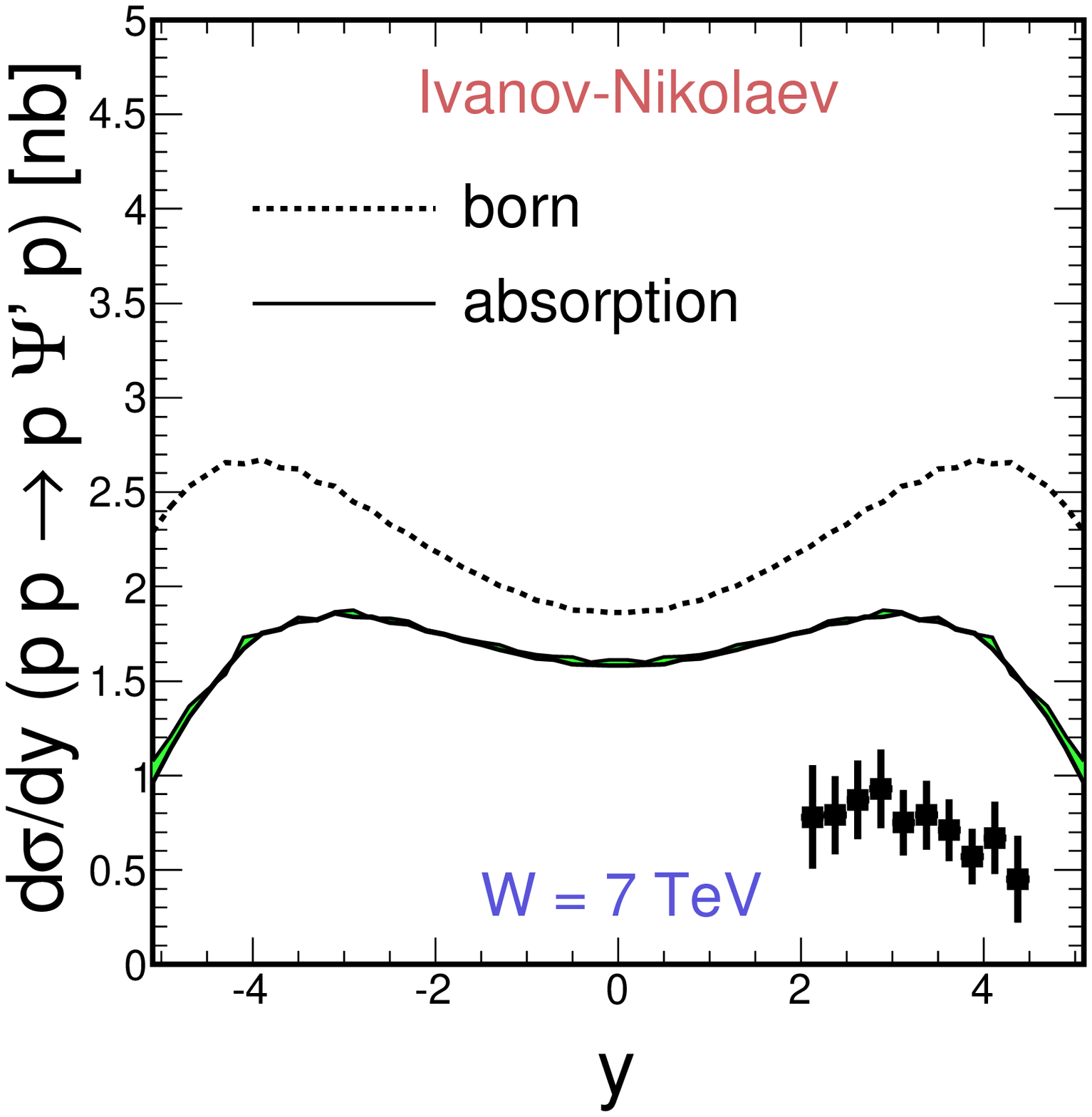}
\includegraphics[width=4.8cm]{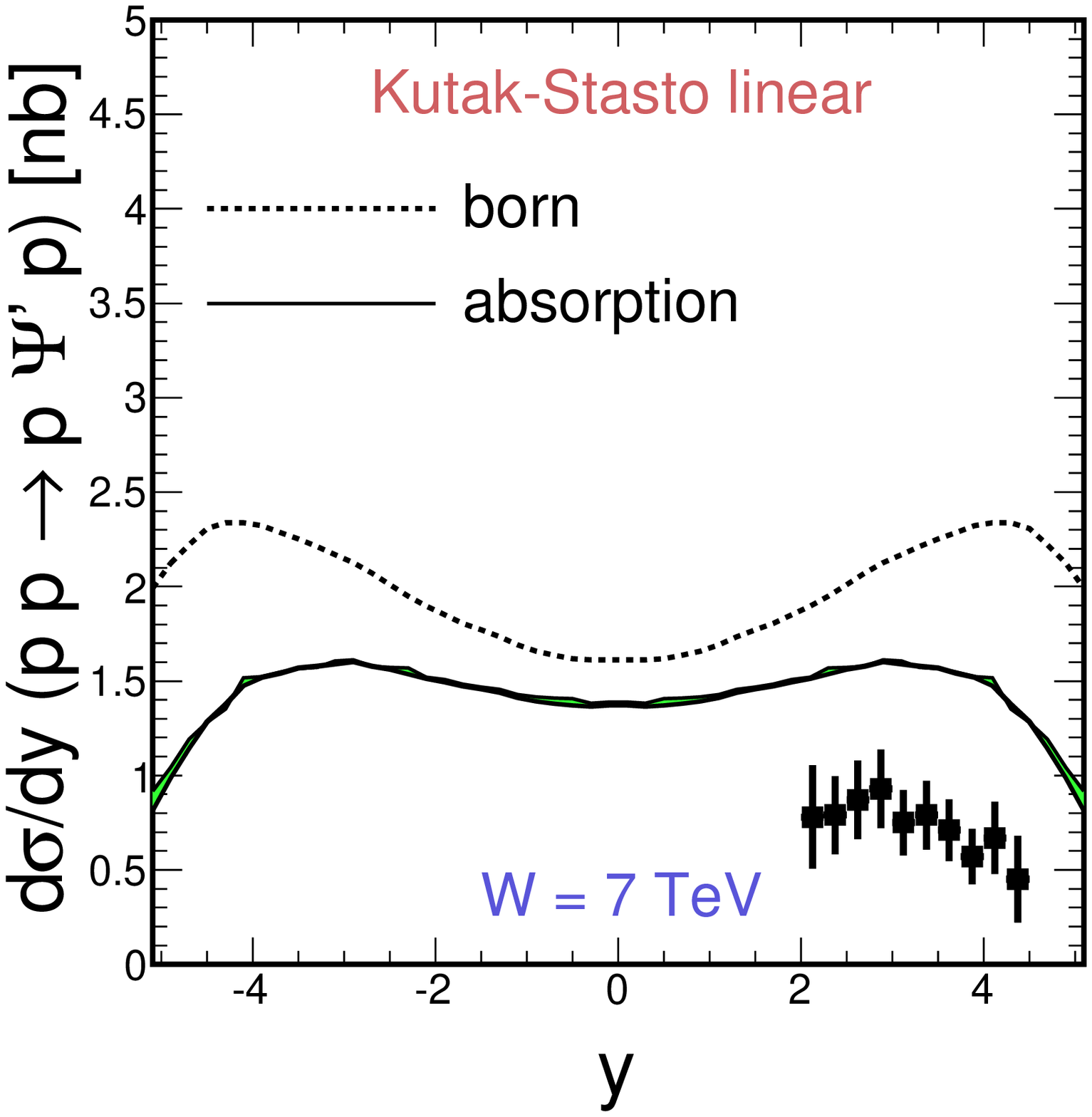}
\includegraphics[width=4.8cm]{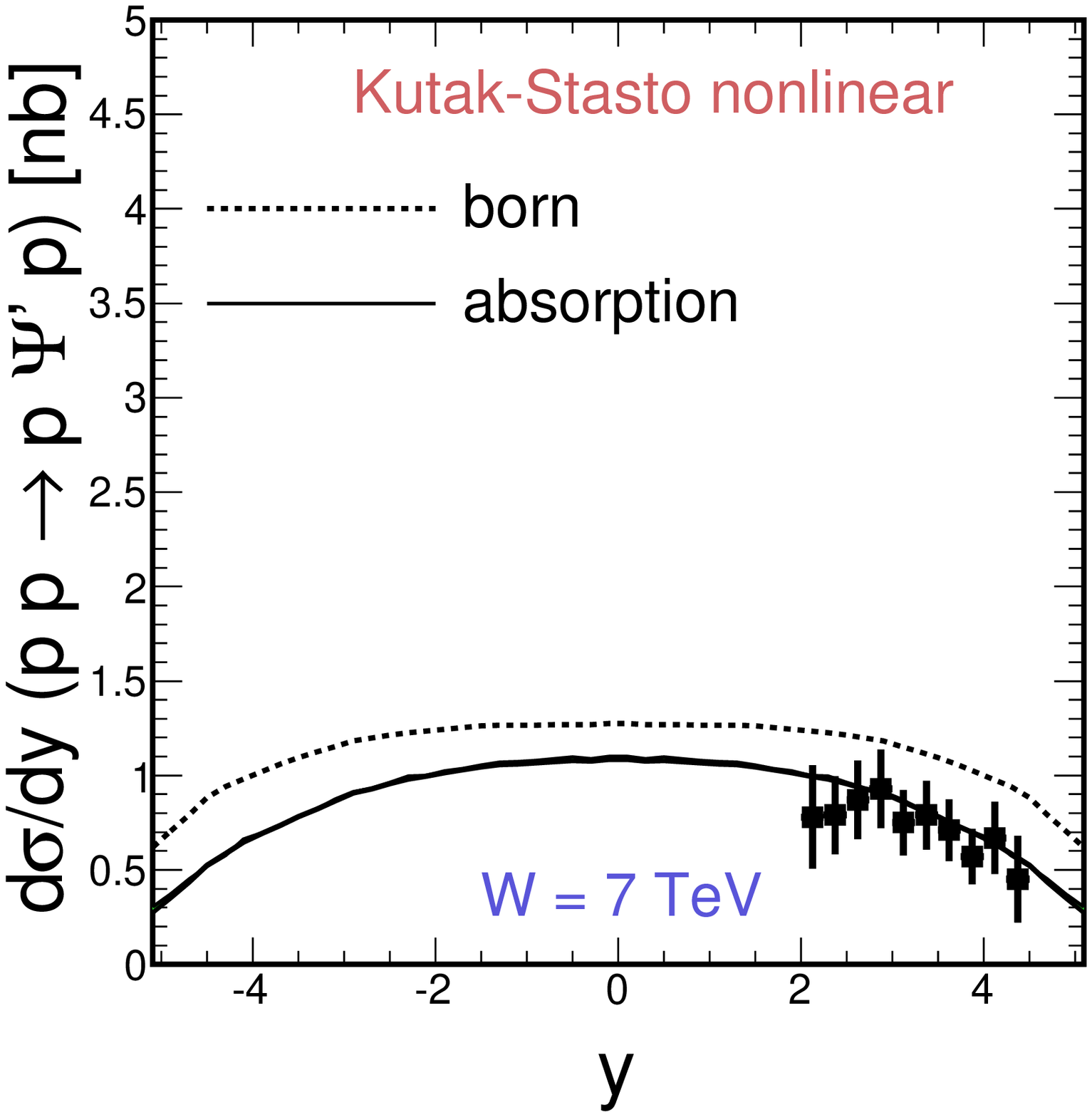}
\caption{ 
$\psi(2S)$ rapidity distribution calculated with inclusion of absorption effects (solid lines), 
compared with the Born result (dashed line) for $\sqrt{s}$ = 7 TeV. 
The LHCb data points \cite{Aaij:2014iea} are shown for comparison.
}
\label{fig:3}
\end{figure}

The exclusive photoproduction of $J/\psi$ and $\psi'(2S)$ charmonia has first been observed 
by the CDF collaborationa at the Tevatron \cite{Aaltonen:2009kg} and recently been
measured by the LHCb collaboration \cite{Aaij:2013jxj,Aaij:2014iea} at the LHC. 
In Fig. \ref{fig:2},
we compare our calculations of the $J/\psi$ rapidity distributions to the LHCb data, in
Fig. \ref{fig:3} we show the comparison with data for the case of $\psi(2S)$ production.

\begin{figure}
\includegraphics[width=.5\textwidth]{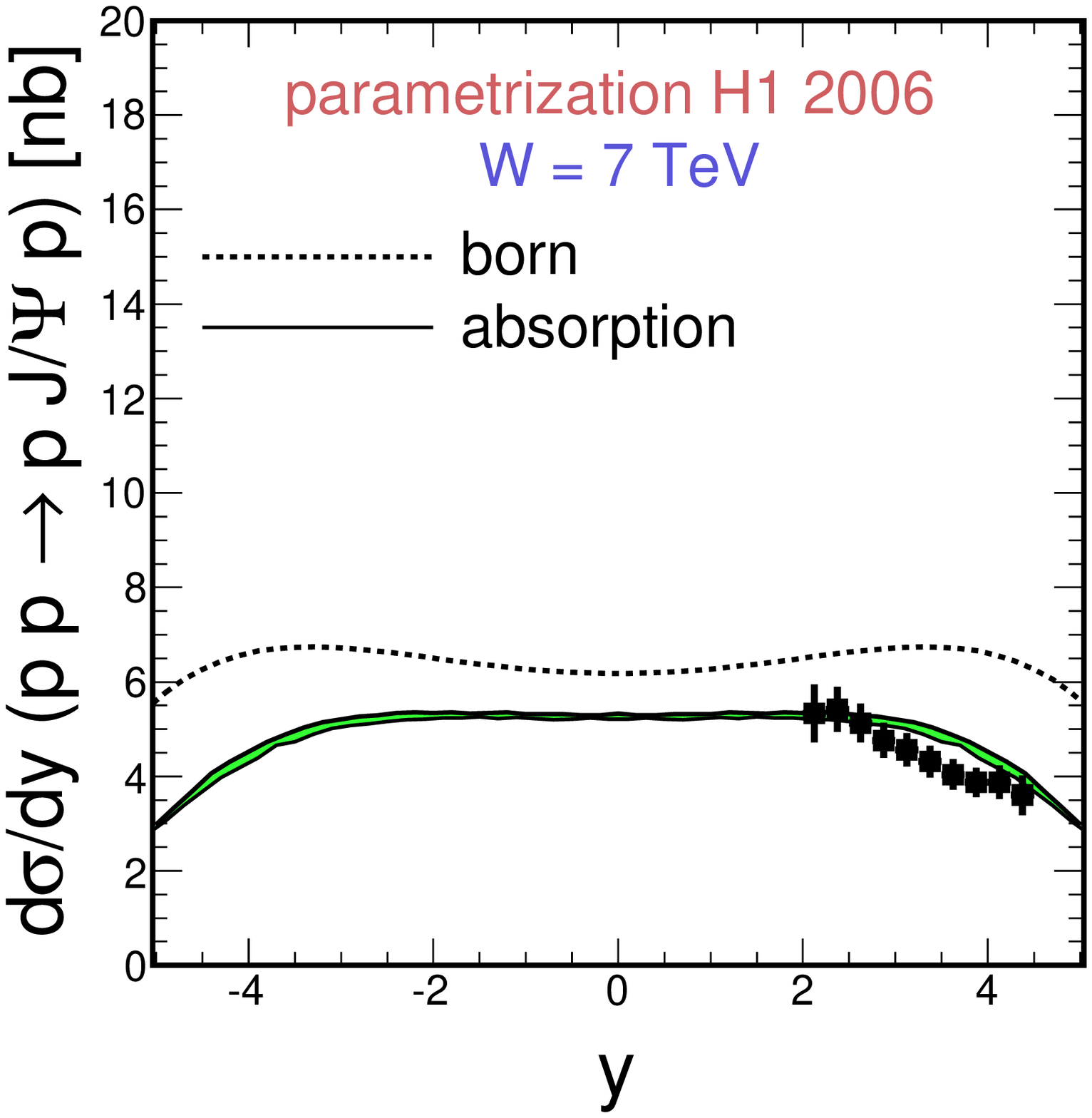}
\includegraphics[width=.5\textwidth]{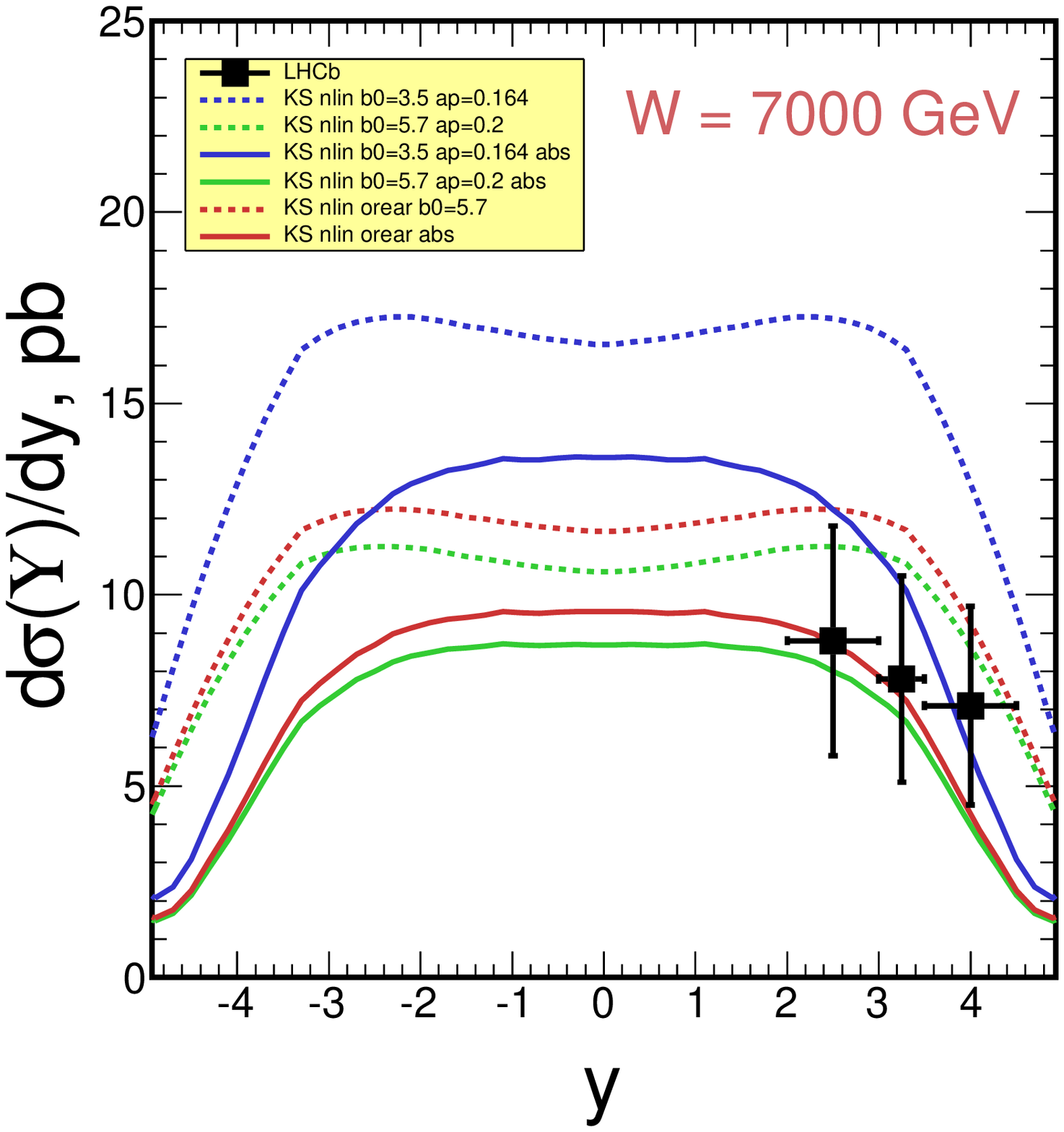}
\caption{Left panel: $J/\psi$ rapidity distribution calculated with a parametrization of the photoproduction amplitude previously used in
\cite{Schafer:2007mm}. Right panel: Rapidity distribution of exclusive $\Upsilon$'s. Data are taken from \cite{Aaij:2015kea}.  
}
\label{fig:4}
\end{figure}

In the leftmost panels we show the result using the Ivanov-Nikolaev glue \cite{Ivanov:2000cm}. 
While it gives a reasonable account of the shape of the rapidity distribution, it overestimates the 
cross section.  A similar job is done by a gluon distribution of \cite{Kutak:2004ym}, 
which solves a BFKL-type linear evolution equation.
A quite good agreement is found for the unintegrated glue of Kutak and Sta\'sto \cite{Kutak:2004ym}
from the nonlinear evolution equation (rightmost panels). 
We stress that all the unintegrated gluon distributions reproduce the Tevatron data \cite{Aaltonen:2009kg},
see \cite{Cisek:2014ala}.
Somewhat ironically also a straight-line extrapolation
of a H1-fit used previously in \cite{Schafer:2007mm} gives a good fit of the cross section
(see the left panel of Fig. \ref{fig:4}).
As it would correspond to the exchange of a single --effective-- Pomeron trajectory, it casts 
a certain doubt on the necessity of saturation effects in the unintegrated glue.
Finally, in the right panel of Fig. \ref{fig:4} we show a comparison of our results for $\Upsilon$ production
compared to recent LHCb data \cite{Aaij:2015kea}. 

\section{Diffractive photoproduction with electromagnetic dissociation}

\begin{figure}
\vspace{1cm}
\includegraphics[width = .5 \textwidth]{dsigdmx.eps}
\includegraphics[width = .5 \textwidth]{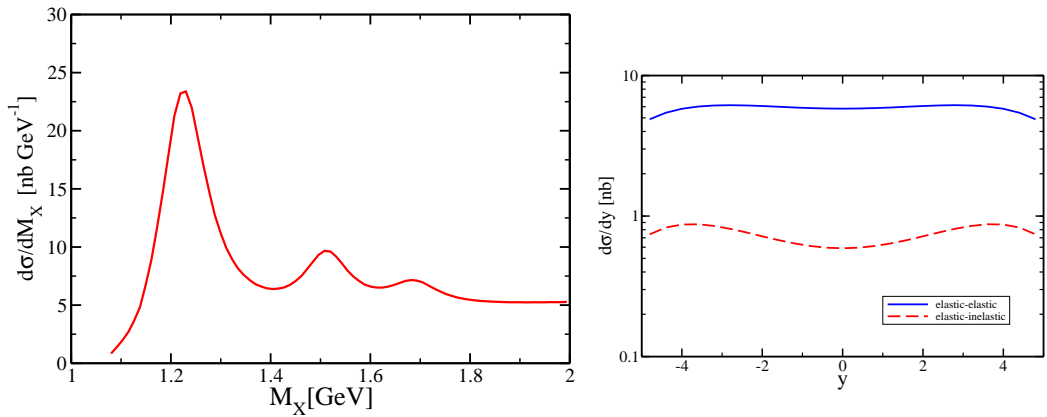}
\caption{
Left panel: distribution in the mass of the electromegnetically excite system, $M_X$ for the $ p p \to X J/\psi p$ process.
Right panel: Rapidity distribution of $J/\psi$ for the same process, for $M_X < 2$ GeV.
}
\label{fig:5}
\end{figure}

Proton dissociative events are of great practical importance as an often 
difficult background to exclusive events. They are also of general interest 
for our understanding of diffractive scattering and Pomeron physics.
Here we concentrate on a simple example relevant to exclusive vector meson production:
the events with electromagnetic dissociation of one of the protons.
The cross section for such processes can be written as:
\begin{eqnarray}
 {d \sigma (pp \to X V p; s) \over dy d^2\bp} = 
  \int {d^2\bq \over \pi \bq^2} {\cal{F}}^{(\mathrm{in})}_{\gamma/p}(z_+,\bq^2) 
  {1\over \pi} {d \sigma^{\gamma^* p \to Vp} \over dt}(z_+s,t = -(\bq - \bp)^2) +( z_+ \leftrightarrow z_-),  
\end{eqnarray}
where $z_\pm = e^{\pm y} \sqrt{(\bp^2 + m_V^2)/s}$. Here the flux of photons associated with the breakup of the proton
is calculable in terms of the structure function $F_2$ of a proton \cite{daSilveira:2014jla,Luszczak:2015aoa}: 
\begin{eqnarray}
 {\cal{F}}^{(\mathrm{inel})}_{\gamma/p}(z,\bq^2) = {\alpha_{\mathrm{em}} \over \pi} (1 - z) \int^\infty_{M^2_{\mathrm{thr}}} 
 {dM_X^2 F_2(x_{Bj},Q^2)  \over M_X^2 + Q^2 - m_p^2}  \Big[ {\bq^2 \over \bq^2 + z (M_X^2 - m_p^2) + z^2 m_p^2} \Big]^2 
\, ,
\end{eqnarray}
where
\begin{eqnarray}
Q^2 =  {1 \over 1 - z} \Big[ \bq^2 + z (M_X^2 -m_p^2)  + z^2 m_p^2 \Big], x_{Bj} = {Q^2 \over Q^2 + M_X^2 - m_p^2} .
\end{eqnarray}
We are especially interested in the region of small invariant masses, $M_X < 2$ GeV, where resonance
excitation is important. Our recent study \cite{Luszczak:2015aoa} showed, that here it is best to use a parametrization
of $F_2$ given in \cite{Fiore:2002re}.
In the left panel of Fig. \ref{fig:5} we show a distribution 
in invariant mass of the excited system for the $pp \to X J/\psi p$ reaction.
In the right panel of Fig. \ref{fig:5} we show the rapidity distribution of $J/\psi$ for elastic and 
e.m. dissociative events. The e.m. dissociation can reach $10 \div 15$\% of the exclusive cross section.

\section{Conclusions}

\begin{itemize}

\item We have compared our $k_\perp$-factorization results with recent 
LHCb ($p p \longrightarrow p\ V \ p$) data, for $V = J/\psi,\psi(2S), \Upsilon$.
The best description is obtained for a glue which does contain saturation effects.

\item Our calculations \cite{Cisek:2014ala} also demonstrate, that absorptive corrections are a strong function of kinematics. 
At large $p_T$, relevant for Odderon searches, 
the Pauli coupling needs to be included. There is a sizeable uncertainty due to absorption in the $p_T$ distribution.

\item Proton dissociation is a background to exclusive processes. Electromagnetic dissociation 
is calculable from $F_2$, excited states $M_X < 2 \, \rm{GeV}$ make a contribution of $10 \div 15 \%$ of the 
exclusive cross section for $J/\psi$.

\end{itemize}


\begin{thebibliography}{99}

\bibitem{Klein:2003vd}
  S.~R.~Klein and J.~Nystrand,
  Phys.\ Rev.\ Lett.\  {\bf 92} (2004) 142003
  [hep-ph/0311164].

\bibitem{Ivanov:2004ax}
  I.~P.~Ivanov, N.~N.~Nikolaev and A.~A.~Savin,
  Phys.\ Part.\ Nucl.\  {\bf 37} (2006) 1
  [hep-ph/0501034].

\bibitem{Schafer:2007mm}
  W.~Sch\"afer and A.~Szczurek,
  Phys.\ Rev.\ D {\bf 76} (2007) 094014
  [arXiv:0705.2887 [hep-ph]].
  
\bibitem{Cisek:2014ala}
  A.~Cisek, W.~Sch\"afer and A.~Szczurek,
  JHEP {\bf 1504} (2015) 159
  [arXiv:1405.2253 [hep-ph]].
  
\bibitem{Chekanov:2002xi}
  S.~Chekanov {\it et al.} [ZEUS Collaboration],
  Eur.\ Phys.\ J.\ C {\bf 24} (2002) 345
  [hep-ex/0201043].

\bibitem{Aaltonen:2009kg}
  T.~Aaltonen {\it et al.} [CDF Collaboration],
  Phys.\ Rev.\ Lett.\  {\bf 102} (2009) 242001
  [arXiv:0902.1271 [hep-ex]].
  
\bibitem{Aaij:2013jxj}
  R.~Aaij {\it et al.} [LHCb Collaboration],
  J.\ Phys.\ G {\bf 40} (2013) 045001
  [arXiv:1301.7084 [hep-ex]].
  
\bibitem{Aaij:2014iea}
  R.~Aaij {\it et al.} [LHCb Collaboration],
  J.\ Phys.\ G {\bf 41} (2014) 055002
  [arXiv:1401.3288 [hep-ex]].
 
\bibitem{Ivanov:2000cm}
  I.~P.~Ivanov and N.~N.~Nikolaev,
  Phys.\ Rev.\ D {\bf 65} (2002) 054004
  [hep-ph/0004206].
  
\bibitem{Kutak:2004ym}
  K.~Kutak and A.~M.~Sta\'sto,
  Eur.\ Phys.\ J.\ C {\bf 41} (2005) 343
  [hep-ph/0408117].
 
\bibitem{Aaij:2015kea}
  R.~Aaij {\it et al.} [LHCb Collaboration],
  JHEP {\bf 1509} (2015) 084
  [arXiv:1505.08139 [hep-ex]].
  
\bibitem{daSilveira:2014jla}
  G.~G.~da Silveira, L.~Forthomme, K.~Piotrzkowski, W.~Sch\"afer and A.~Szczurek,
  JHEP {\bf 1502} (2015) 159
  [arXiv:1409.1541 [hep-ph]].
 
\bibitem{Luszczak:2015aoa}
  M.~{\L}uszczak, W.~Sch\"afer and A.~Szczurek,
  Phys.\ Rev.\ D {\bf 93} (2016) no.7,  074018
  [arXiv:1510.00294 [hep-ph]].

\bibitem{Fiore:2002re}
  R.~Fiore, A.~Flachi, L.~L.~Jenkovszky, A.~I.~Lengyel and V.~K.~Magas,
  Eur.\ Phys.\ J.\ A {\bf 15} (2002) 505
  [hep-ph/0206027].

\end{thebibliography}
\end{document}